\documentclass[floatfix,showpacs,preprintnumbers,amsmath,amssymb,aps,twocolumn,prl,10pt]{revtex4}

\usepackage{amsfonts}
\usepackage{amsmath}
\usepackage[pdftex]{graphicx}
\usepackage[pdftex,usenames]{color}
\usepackage{bm}
\usepackage{multirow}

\begin{document}

\title{Coupled Ferromagnetic and Nematic Ordering of Fermions in an Optical Flux Lattice}
\author{Stefan K. Baur}
\affiliation{T.C.M. Group, Cavendish Laboratory, J. J. Thomson Avenue, Cambridge CB3 0HE, United Kingdom}
\author{Nigel R. Cooper}
\affiliation{T.C.M. Group, Cavendish Laboratory, J. J. Thomson Avenue, Cambridge CB3 0HE, United Kingdom}

\date{\today}

\begin{abstract}
Ultracold atoms in Raman-dressed optical lattices allow for effective momentum-dependent interactions among single-species fermions originating from short-range $s$-wave interactions. These dressed-state interactions combined with very flat bands encountered in the recently introduced optical flux lattices push the Stoner instability towards weaker repulsive interactions, making it accessible with current experiments. As a consequence of the coupling between spin and orbital degrees of freedom, the magnetic phase features Ising nematic order.
\end{abstract}
\pacs{67.85.Lm, 03.75.Ss, 03.65.Vf, 73.22.Gk}
\maketitle

Recently, considerable effort has been made to observe the Stoner
instability to itinerant ferromagnetism with ultracold
gases~\cite{Duine:2005lh, Conduit:2009qf, Jo:2009fu, Pilati:2010nx,
  Chang:2010cr}. So far, this effort has been fruitless and it has
been argued that rapid dimer formation at the large coupling
strength that is required for ferromagnetism precludes the formation
of magnetic domains~\cite{Pekker:2011oq, Sanner:2011kl}. Here we show
that atoms subjected to optical lattices involving coherent Raman coupling
of internal states can have a strongly enhanced Stoner instability. The ferromagnetic phase
appears at much weaker
coupling strength where the gas is less susceptible to
  dimer
  formation. Furthermore, our results display several intriguing novel phenomena, such as interaction-induced phase transitions between distinct Fermi surface topologies and nematic ordering,  allowing close parallels between the physics of cold gases and phenomena in diverse systems such as high-temperature superconductors, ruthenates and quantum Hall systems~\cite{Fradkin:2010hc}.

  Central to our studies are the novel effects that arise when atoms
  are subjected to Raman dressing.  Raman dressing has recently been
  used in experiments to create artificial gauge
  potentials~\cite{Lin:2009qf,Wang:2012ve,zwierlein} and to induce effective higher
  partial-wave interactions among identical bosons from short-ranged
  $s$-wave interactions~\cite{Williams:2012kx}. We consider an atomic
  Fermi gas subjected to an optical flux lattice~\cite{Cooper:2011kl},
  in which both of these effects are important.  The orbital effects
  of the gauge field cause the lowest energy band of the optical flux
  lattice to be very narrow in energy even for a shallow lattice far
  from the tight-binding limit. Interactions among fermionic atoms in
  this lowest band remain sizeable~\cite{Cooper:2011cr}.  Simple
  s-wave interactions between distinct bare fermions give rise to
  effective interactions of nonzero range among single species fermions of
  the lowest band~\cite{Cooper:2011cr,Cui:2012tg}. We show that
    these interactions, within this narrow band, cause a
  ferromagnetic transition at a much smaller coupling than that in the
  continuum. In view of the coupling of spin and orbital motion
    through the Raman dressing, the ferromagnetic transition appears
    as a change in Fermi surface topology.  Furthermore, it is
  accompanied by a reduction of the (spatial) crystal symmetry so
 also involves nematic order~\cite{Fradkin:2010hc}. As we describe, this coupling of spin and orbital
    motion allows the magnetic/nematic ordering to be readily
    measured in experiment by band-mapping techniques.

\begin{figure}
\includegraphics[width= \columnwidth]{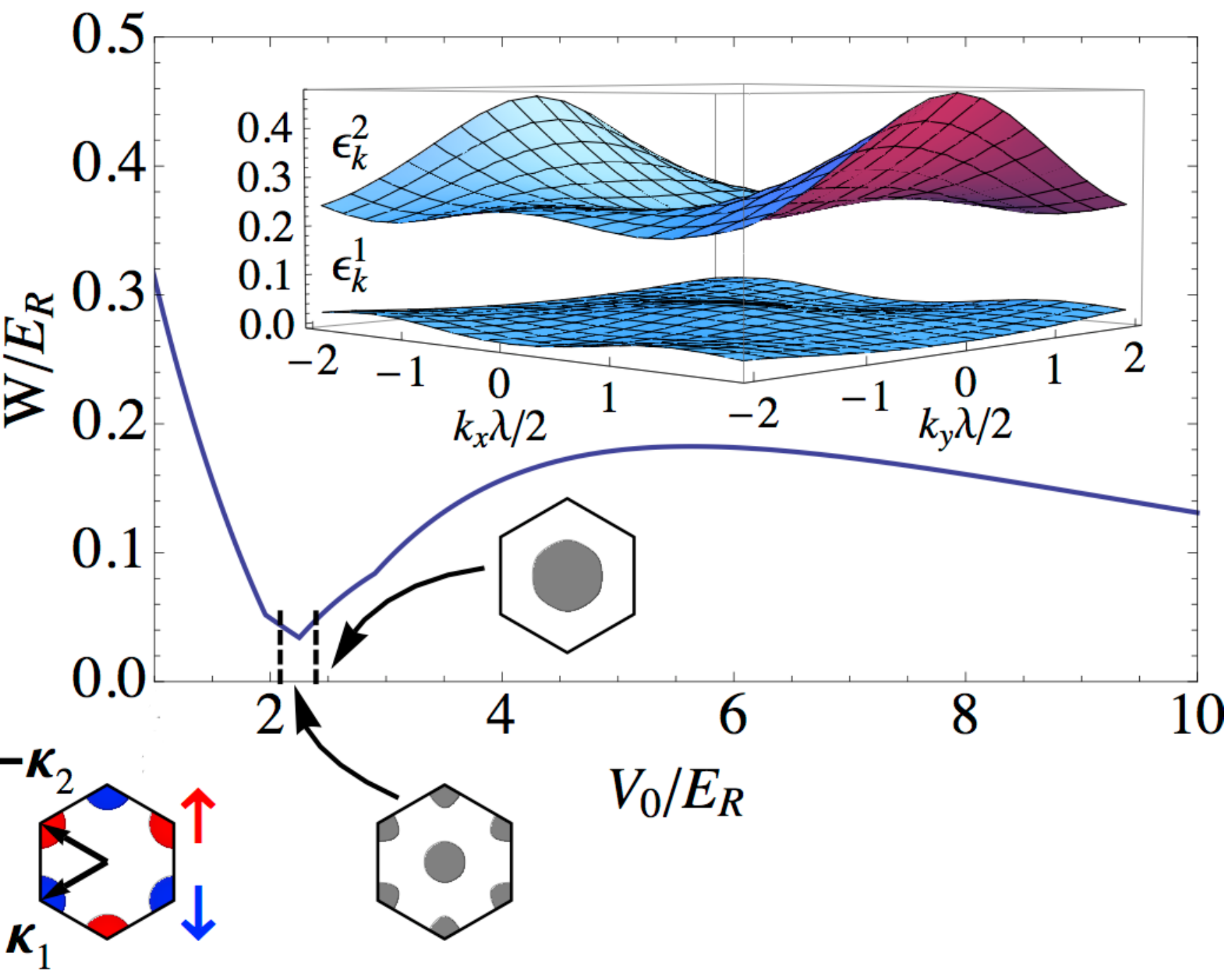}
\caption{(Color online) Bandwidth $W$ of the lowest band as a function
  of lattice depth of the optical flux lattice discussed in the main
  text (with $\theta=0.3$ and $\epsilon=0.4$). The inset shows the
  dispersion of the lowest two bands for $V_0/E_R=2.25$. As the
  lattice is ramped up, the dispersion of the lowest band develops a
  minimum at the center of the Brillouin zone causing reconstruction
  of the Fermi surface for noninteracting particles. The locations
  where these reconstructions occur for filling $\nu=1/4$ are marked
  by the dashed lines. For
    vanishing lattice depth $V_0=0$ the Fermi circles of the
    spin-$\uparrow$ (red) and spin-$\downarrow$ (blue) are displaced
    from each other to the corners of the Brillouin zone.}
  \label{fig:narrowbands}
\end{figure}

We consider the implementation of an optical flux lattice of
  Ref.~\cite{Cooper:2011cr}, which involves two-photon dressing of
  hyperfine states. We focus on an atomic species with ground-state angular momentum $F=1/2$, but
  our key ideas are readily extended to atoms with larger $F$ as described at the
  end of the Letter.  We further restrict attention to a quasi-2D
  geometry, assuming an in-plane confinement energy that is large compared to
  all other energy scales.

The optical flux lattice is formed by interference of three linearly
polarized in-plane laser beams with wave vectors
$\bm{\kappa}_1=-\kappa/2(\sqrt{3},1)$, $\bm{\kappa}_2=\kappa/2(\sqrt{3},-1)$,
and $\bm{\kappa}_3=\kappa (0,1)$. A fourth circular polarized laser
beam, oriented perpendicular to the 2D plane, provides the other frequency required for two-photon Raman coupling of the Zeeman-split hyperfine states.

This optical potential leads to a single particle Hamiltonian
\begin{eqnarray}
\label{eq:gaugepotential}
\hat{H}_0 = \frac{\mathbf{p}^2}{2m} \hat{\openone} +  V_{\rm sc}(\mathbf{r}) \hat{\openone}+ \hat{{\bm \sigma}} \cdot \mathbf{B} (\mathbf{r}),
\end{eqnarray}
in which the atom experiences a scalar potential
\begin{eqnarray} 
V_{\rm sc}= 
V_0 (3 \cos^2(\theta)-1) \sum_j \cos({\bm \kappa}_{j}'  \cdot \mathbf{r})
\end{eqnarray}
and its spin $\hat{{\bm \sigma}}$ couples to an effective magnetic field
\begin{eqnarray}
B_z
=\sqrt{3}\, V_0 \sin^2(\theta) \sum_j \sin({\bm \kappa}_j' \cdot \mathbf{r})\\
B_x
+i B_y
=\epsilon V_0  \cos(\theta) \sum_j e^{-i {\bm \kappa}_j \cdot \mathbf{r}}.
\end{eqnarray}
$V_0$ denotes the lattice depth, $\theta$ is the polarization angle of
the in-plane beams with respect to the surface normal, $\epsilon$ is
proportional to the ratio between Raman coupling and scalar potential,
and $\bm{\kappa}_1'=\bm{\kappa}_1-\bm{\kappa}_2$,
$\bm{\kappa}_2'=\bm{\kappa}_3-\bm{\kappa}_1$, and
$\bm{\kappa}_3'=\bm{\kappa}_2-\bm{\kappa}_3$~\cite{Cooper:2011cr}.
The geometry of the Raman beams is such that conversion from
spin-$\uparrow$ to spin-$\downarrow$ involves a momentum exchange of
$\bm{\kappa}_1$, $\bm{\kappa}_2$ or $\bm{\kappa}_3$. This causes the
spin character of the Bloch states to vary with crystal momentum
within the Brillouin zone. (The reciprocal lattice
basis vectors can be taken to be $\mathbf{G}_1 \equiv \bm{\kappa}_1 -
\bm{\kappa}_3$ and $\mathbf{G}_2 \equiv \bm{\kappa}_2 -
\bm{\kappa}_3$.)  Notably, for vanishing lattice depth $V_0=0$,
when the energy eigenstates are simply plane waves for spin-$\uparrow$
and spin-$\downarrow$, the crystal momentum for the spin-$\downarrow$
($\uparrow$) state with zero kinetic energy is simply
$\bm{\kappa}_{1}$ ($-\bm{\kappa}_{2}$), or any equivalent point
  related by the addition of reciprocal lattice vectors. Hence,
  the crystal momenta of the two spin states are displaced from each
  other within the Brillouin zone.  In this limit $V_0=0$, an
  unpolarized state of noninteracting fermions therefore appears as
  two filled Fermi circles, centered on $\bm{\kappa}_{1}$
  ($-\bm{\kappa}_{2}$) for spin-$\downarrow$ ($\uparrow$) and shown in
  blue (red) in Fig.~\ref{fig:narrowbands}. The difference from the 
  conventional picture of two Fermi circles centered on $\mathbf{k}=0$ 
  just reflects the spin-dependent momentum offsets common to all forms
  of Raman coupling involving momentum exchange~\cite{Lin:2009qf}.

  The width of the lowest-energy band is shown in
  Fig.~\ref{fig:narrowbands} as a function of the overall lattice
  depth $V_0$ for fixed values of $\epsilon$ and $\theta$. The
  bandwidth passes through a minimum at $V_0/E_R \simeq 2$, with the recoil energy defined by $E_R \equiv \hbar^2\kappa^2/(2m)$. This
    is the regime where the optical flux lattice best mimics the
    orbital effects of a uniform magnetic field. The lowest energy band is
    similar to a Landau level: with small bandwidth and Chern number
    of one~\cite{Cooper:2011cr}. In the vicinity of this point the positions of the band
  minima change within the Brillouin zone. (Similar features are found
  in tight-binding models of Chern insulators when next
  nearest-neighbor hoppings are included~\cite{Tang:2011il}.) This
  reconstruction is illustrated in Fig.~\ref{fig:narrowbands} by the
  non-interacting Fermi surfaces shown for a band filling of
  $\nu=1/4$.  Note that at $V_0 =0$ the Fermi surface consists of two
  disconnected circles: these are the spin-up and spin-down Fermi
  surfaces, displaced in crystal momentum as described
  above.

For nonzero $V_0/E_R$ the 
  spin composition of the Bloch state continuously varies with crystal momentum.
 As a result, $s$-wave interactions between spin-up and spin-down
components lead to effective momentum-dependent interactions between fermions
in this band. It is remarkable that even though we started with a model of short-ranged interactions, we obtain an effective theory of interacting spinless fermions~\cite{Cooper:2011cr}. 

Atoms restricted to states in the lowest band are described by the effective Hamiltonian
\begin{eqnarray}
\label{eq:hlb}
H_{lb}=\sum_{\mathbf{k}} \epsilon_\mathbf{k} c_{\mathbf{k}}^{\dagger} c_{\mathbf{k}}+\frac{1}{2}\sum_{\mathbf{k}_1 \mathbf{k}_2  \mathbf{k}_3 \mathbf{k}_4} V_{\mathbf{k}_1 \mathbf{k}_2 \mathbf{k}_3 \mathbf{k}_4} c_{\mathbf{k}_1}^{\dagger} c_{\mathbf{k}_2}^{\dagger} c_{\mathbf{k}_3} c_{\mathbf{k}_4}
\end{eqnarray}
where $\epsilon_{\mathbf{k}}$ is the band dispersion, 
$c_{\mathbf{k}}^{(\dagger)}$ are the fermionic field operators for state of crystal momentum $\mathbf{k}$, and

\begin{equation}
\nonumber
V_{\mathbf{k}_1 \mathbf{k}_2 \mathbf{k}_3 \mathbf{k}_4}=g_{\rm 2D} \int d^2\mathbf{r}\; \sum_{\sigma}\phi_{\mathbf{k}_1 \sigma}^*(\mathbf{r})  \phi_{\mathbf{k}_2 \bar{\sigma}}^*(\mathbf{r})  \phi_{\mathbf{k}_3 \bar{\sigma}}(\mathbf{r})   \phi_{\mathbf{k}_4 \sigma}(\mathbf{r})
\end{equation}
is the effective interaction in the lowest band in terms of the eigenfunctions $(\phi_{\mathbf{k} \uparrow}(\mathbf{r}),\phi_{\mathbf{k} \downarrow}(\mathbf{r}))^T$ of the single particle Hamiltonian (\ref{eq:gaugepotential}). In the following we characterize the bare interaction strength by the dimensionless coupling parameter $\tilde{g} \equiv  mg_{\rm 2D}/\hbar^2$. In terms of the 3D $s$-wave scattering length $a_s$ and the harmonic oscillator length of the transverse confinement $l_z$ (assuming the atoms are confined to a 2D plane via a tight harmonic potential along the $z$ axis) one has $\tilde{g}=\sqrt{8 \pi} a_s/l_z$, valid in the limit where $|a_s| \ll l_z$~\cite{footnote2}.

To study the effects of interactions, we perform a Hartree-Fock (HF) variational approximation which results in the energy functional
\begin{eqnarray}
\label{eq:ehf}
E[\{n_\mathbf{k}\}]=\sum_{\mathbf{k}} \epsilon_{\mathbf{k}} n_{\mathbf{k}}+\frac{1}{2} \sum_{\mathbf{k} \mathbf{k}'} V_{\mathbf{k} \mathbf{k}'} n_{\mathbf{k}} n_{\mathbf{k}'}
\end{eqnarray}
with
$V_{\mathbf{k}\mathbf{k}'}=V_{\mathbf{k} \mathbf{k}'\mathbf{k}' \mathbf{k}}-V_{\mathbf{k} \mathbf{k}' \mathbf{k} \mathbf{k}'}$.
For our zero-temperature results, we find the ground states that minimize Eq.~\eqref{eq:ehf} 
for a fixed total number of particles $N=\sum_{\mathbf{k}} n_\mathbf{k}$. This is achieved by setting the occupation numbers $n_{\mathbf{k}}$ equal to unity for the $N$ orbitals with lowest
HF energies 
\begin{eqnarray}
\label{eq:energies}
\xi_{\mathbf{k}}&=&\epsilon_{\mathbf{k}}+\sum_{\mathbf{k}'} V_{\mathbf{k} \mathbf{k}'} n_{\mathbf{k}'} \,.
\end{eqnarray}
We determine these energies self-consistently by numerical iteration,
discretizing momenta of the Brillouin zone on a fine grid~\cite{footnote}.  For our results at nonzero temperature $T$,
we instead find $n_{\mathbf{k}}$ by minimizing the thermodynamic
potential $\Omega=E-TS-\mu N$ where $\mu$ is the chemical potential
and $S=-k_B \sum_{\mathbf{k}} n_{\mathbf{k}}
\ln(n_{\mathbf{k}})+(1-n_{\mathbf{k}}) \ln(1-n_{\mathbf{k}})$ is the
entropy. The grand canonical potential is stationary 
($\delta
\Omega/\delta n_{\mathbf{k}}=0$) when $n_{\mathbf{k}}=1/(e^{(\xi_{\mathbf{k}}-\mu)/k_B T}+1)$
which we numerically iterate to self-consistency with Eq.~\eqref{eq:energies}.
We start from trial states with different symmetry and random initial occupation numbers
and then compare their grand canonical potentials in order to find the
minimum free energy configuration~\cite{footnote3}.

Our results show a robust ferromagnetic phase for a wide range
  of parameters. To characterize this phase we use the magnetization
per particle along the $z$ direction as an order parameter
\begin{eqnarray}
m_z=\frac{1}{\nu} \int_{A_{\text{cell}}}  \hspace{-3mm} d^2r\; \left[ n_{\uparrow}(\mathbf{r})-n_{\downarrow}(\mathbf{r}) \right] \,. 
\end{eqnarray}

In Fig.~\ref{fig:phasediagramzerotemperature} we show the
  (modulus of the) calculated order parameter at zero temperature for
  a band filling of $\nu=1/4$, as a function of 2D coupling
  $\tilde{g}$ and of lattice depth $V_0/E_R$. These results are
  representative of other fillings with $\nu \lesssim 1/2$.  In the absence of
  any lattice, $V_0/E_R=0$, there is a transition from paramagnet to
  ferromagnet at ${\tilde g}_c= 2\pi$. This is the conventional
  Stoner instability for fermions with contact repulsion in
  2D~\cite{conduit}. This transition appears as a reconstruction of the
  Fermi surface from two Fermi circles to one Fermi circle.  As the
  lattice depth is increased, our results show a steady decrease of
  the coupling at which the ferromagnetic transition occurs.  The
  minimum coupling for ferromagnetism arises for $V_0/E_R \simeq 2$,
  close to the condition for minimum bandwidth,
  Fig.~\ref{fig:narrowbands}. We find that the interaction
strength ${\tilde g}_c$ required is reduced from its free-space
value by a factor of about $1/4$.
This is one of the main results
  of this Letter: the reduction of coupling as compared to the
  free-space case means that ferromagnetism can be achieved without
  requiring as close an approach to a Feshbach resonance. A change
of coupling by a factor of $1/4$ is expected to have a very dramatic
reduction in the rate of dimer formation~\cite{petrov,ketterle,Pricoupenko:2007fk}. 
Given this reduction in required interaction strength, we can estimate the increase in lifetime in a 2D gas from the experimental data of the Cambridge group~\cite{Koehl2012}, where the lifetime of the repulsive Fermi polaron has been measured as a function of interaction strength. A change of coupling by a factor of $1/4$ is expected to increase the lifetime of the upper branch from about $h/E_F$ by 2 decades to 100 $h/E_F$~\cite{Ngampruetikorn:2011zr}. 
\begin{figure}
\includegraphics[width=\columnwidth,trim= 0 0 100 80]{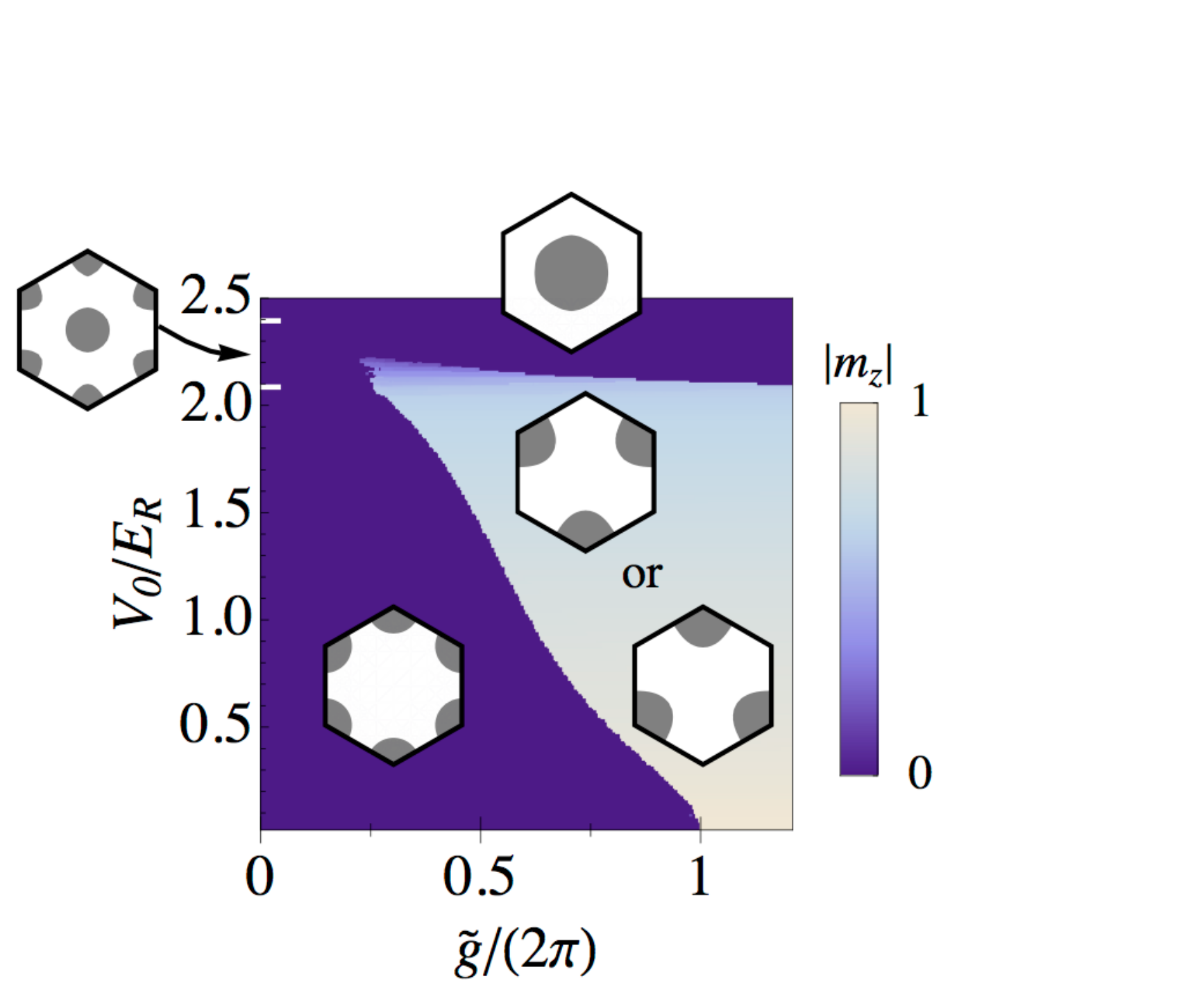}
\caption{Magnetization of the ground state of interacting fermions in the lowest band of the optical flux lattice at uniform filling factor $\nu=1/4$ as a function of lattice depth $V_0/E_R$ and dimensionless coupling strength $\tilde{g}$. The gray shaded areas on the insets illustrate occupied states in the first Brillouin zone. White lines mark transitions between unmagnetized states of different Fermi surface topology.
}\label{fig:phasediagramzerotemperature}
\end{figure}

In addition to transitions between distinct Fermi surface topologies, phases of interacting fermions can also spontaneously break lattice symmetries. For our model, we find that the appearance of ferromagnetism, with nonzero magnetization, is also accompanied by
a  breaking of rotational symmetry. We argue that this is a general feature of Raman-dressed atomic systems.
While spin-rotational invariance is explicitly broken by the coupling of spin and orbital degrees of freedom, the optical flux lattice retains a discrete symmetry. It is invariant under a spin-flip combined with a $2 \pi/6$ rotation in real space
\begin{eqnarray}
\label{eq:sixfold}
\hat{U}_6=\hat{\sigma}_x \hat{R}_{2\pi/6}.
\end{eqnarray}

The Stoner ferromagnetic transition (effectively a Pomeranchuk
instability in the spin channel) causes spontaneous symmetry breaking of the $C_6$
  spin-rotation symmetry arising from $\hat{U}_6$ down to a residual
  $C_3$ symmetry associated with $\hat{U}_6^2 =
  \hat{R}_{2\pi/3}$. This phase transition is analogous to the
lattice symmetry breaking in the electronic Ising nematic phases in
solid-state materials ~\cite{Fradkin:2010hc, Yamase:2000kx, Halboth:2000kx,
  Yamase:2005uq}. Since the order parameter of the symmetry broken
phase is in the 2D Ising universality class, we expect this
phase to survive to nonzero
temperature.

The phase diagram at nonzero temperature is shown in Fig.~\ref{fig:phasediagramfinitetemp} as a function of chemical potential for a lattice with $V_0/E_R = 2$ and $\tilde{g}=1.9$. This shows that the ferromagnetic phase is a robust phase across a range of densities and temperatures. The maximum transition temperature of $k_B T\simeq 0.14 W$ at $\mu-\epsilon_{\rm min} \simeq 1.54 W$ corresponds to an entropy per particle of $S/N \simeq 1.1k_{\rm B} $. Entropies of this order are being reached in current optical lattice experiments~\cite{Jordens:2010vn}.

   While our 
mean-field theory neglects correlations~\cite{Zhai:2009} that can lead to
  quantitative changes in the location of the Stoner instability, detailed 
numerical studies of related models show a robust
  ferromagnetic phase~\cite{Ma:2012,Pilati:2010nx,
  Chang:2010cr}. Strongly
  correlated phases, related to fractional quantum Hall states, which
  cannot be accessed in HF theory, can commonly coexist with 
ferromagnetism (at 
specific fillings)~\cite{dassarmapinczuk,Saiga:2006fk,Katsura:2010}. We expect the reduction of the critical coupling strength 
that we predict to be a robust feature, since the shallow optical flux 
lattice leads to a significant reduction in the bandwidth with small 
decrease in the interaction matrix elements.

Experimental studies of the ferromagnetic transition we predict
  will require the use of an atomic species for which both strong
  interactions and Raman coupling can be achieved without significant
  heating. For a hyperfine ground state of $F=1/2$, the natural
  candidates are $^{171}$Yb or $^{199}$Hg~\cite{Cooper:2011cr}. For
$^{171}$Yb, $s$-wave contact interactions between the two states of the
lowest hyperfine manifold (as spin-$\uparrow$/$\downarrow$) can be
conveniently tuned via an optical Feshbach
resonance~\cite{Ciurylo:2005qa,Reichenbach:2009mi}. Another
  possibility is to use an effective two-level system formed by
  exploiting the quadratic Zeeman effect to Raman couple two of the hyperfine
  states in the $F=9/2$ ground-state manifold of $^{40}$K with
  neighboring $m_F$~\cite{Wang:2012ve}.  The interstate interactions can be conveniently
  tuned by one of the set of magnetic Feshbach resonances that exist
  for these levels.
\begin{figure}
\includegraphics[width=\columnwidth,trim=0 0 40 0]{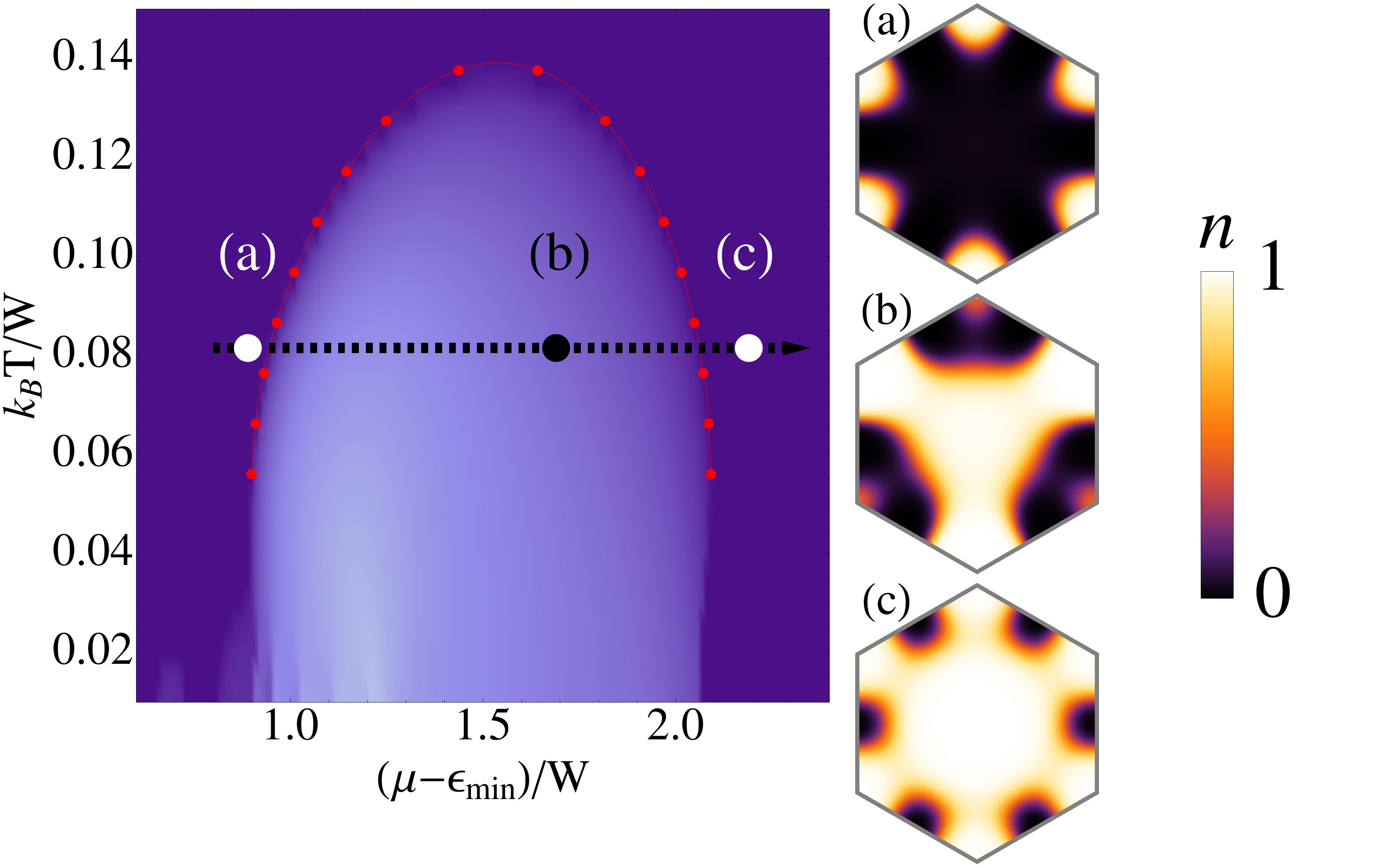}
\caption{Left: Magnetization for $V_0/E_R=2$, $\theta=0.3$, $\epsilon=0.4$ and $\tilde{g}=1.9$ as a function of temperature and chemical potential (scale same for $m_z$ as in Fig. \ref{fig:phasediagramzerotemperature}). $W$ is the bandwidth, and $\mu-\epsilon_{\rm min}$ is the chemical potential measured from the bottom of the lowest band. The red dots lie on the spinodal where the symmetric state becomes unstable (the line is a guide to the eye).  Right: Occupation numbers $n_{\mathbf{k}}$ within the first Brillouin zone for $(\mu-\epsilon_{\rm min})/W=0.89$, $1.69$, $2.19$ correspond to (a), (b), (c).
}\label{fig:phasediagramfinitetemp}
\end{figure}

  The most direct way to measure the order parameter, in
  Figs.~\ref{fig:phasediagramzerotemperature} and
  \ref{fig:phasediagramfinitetemp}, is by individually imaging the
  total spin populations $N_{\uparrow, \downarrow}$. An important
  practical consequence of the coupling of spin and orbital degrees of
  freedom is that total magnetization is not conserved. Hence this
  allows the formation of a macroscopic net magnetization starting
  from an initially unpolarized gas. Measurements of the net
    magnetization will give the average properties of the
    inhomogeneous cloud in the trap.
  If in addition one could measure the local \emph{in situ} spin populations
  $n_{\uparrow, \downarrow}(\mathbf{r})$ along a contour of fixed
  filling factor one could then map out the phase diagram at different
  fillings and lattice depth analogous to recent studies of
  two-dimensional Bose gases~\cite{Hung:2011fu}.
 \begin{figure}
\includegraphics[width=\columnwidth,trim=0 50 0 20]{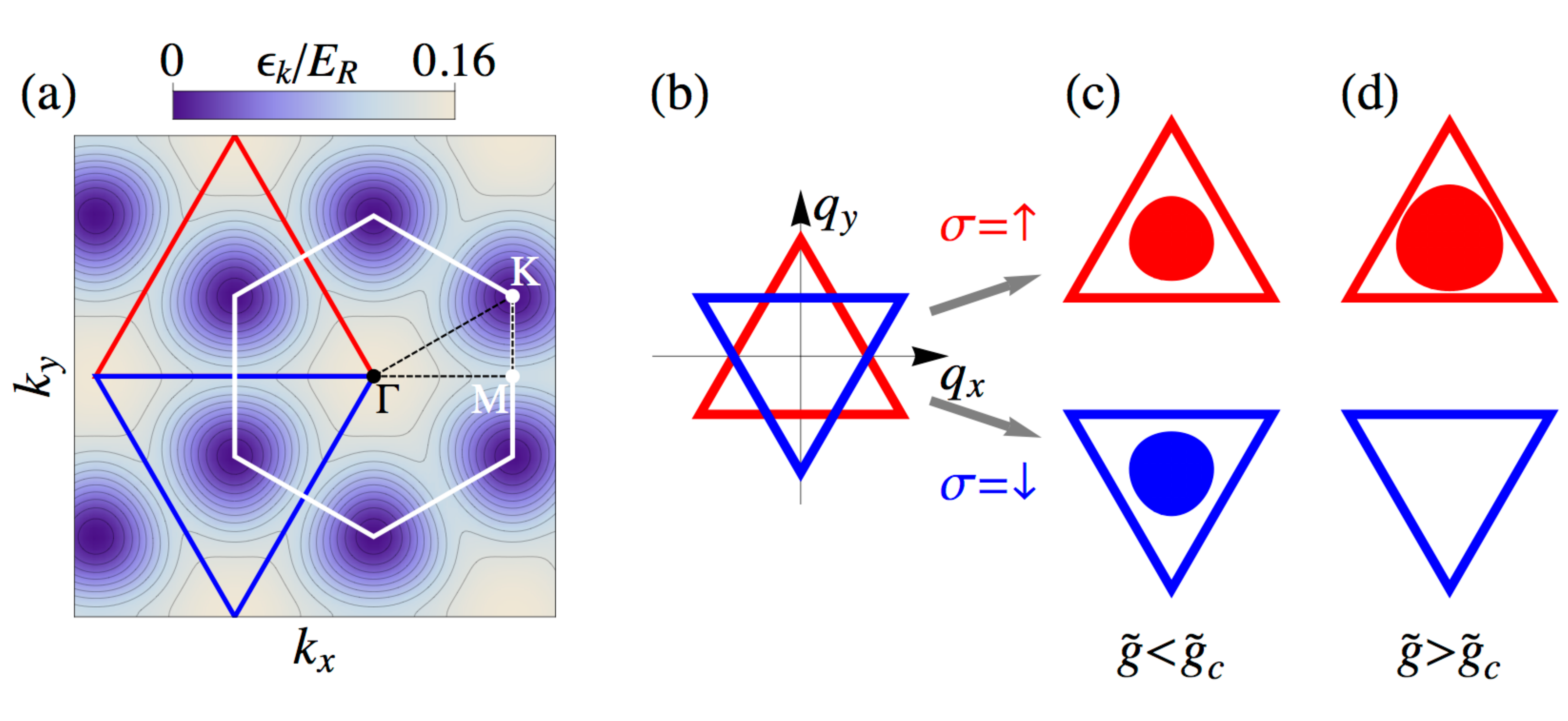}
\caption{Experimental signatures of adiabatic band mapping. (a) Contour plot of the dispersion $\epsilon_{\mathbf{k}}$ for $V_0/E_R=1.8$, indicating a rhombus-shaped reciprocal lattice unit cell. The Bloch states inside the upper red  (lower blue)  triangle 
in (a) map to spin-$\uparrow$ ($\downarrow$) states with free-space momentum $\mathbf{q}$ illustrated in (b).
For a completely filled lowest band one would observe a fully occupied hexagram (b).
When the atoms pass subsequently through a Stern-Gerlach filter, the two spin states can be separately resolved. This would allow clear signatures of the transition from the (c) unmagnetized  to the (d) magnetized phase.
}\label{fig:bandmapping}
\end{figure}
 
 Another complementary probe sensitive to the (trap averaged) Fermi surface of the dressed lowest band fermions is the adiabatic band mapping technique~\cite{Greiner:2001ye,Kohl:2005qo}. This probe has the additional advantage that it also allows the detection of unmagnetized phases with different Fermi surface topologies. Here the lattice potential is ramped down at a rate slow compared to that of the band gap and fast compared to that of many-particle dynamics. Then the Raman-dressed Bloch states are adiabatically mapped onto free-particle plane-wave states of definite spin, which can be imaged after time-of-flight expansion.

   The form of this mapping can be deduced by recalling that for
   vanishing lattice depth, $V_0=0$, the spin-$\downarrow$
   ($\uparrow$) free-particle state with zero kinetic energy has
   crystal momentum $\bm{\kappa}_{1}$ ($-\bm{\kappa}_{2}$) (or any
   point related by reciprocal lattice vectors). Under adiabatic
   band mapping, a Bloch state whose crystal momentum $\mathbf{k}$ is
   closer to $\bm{\kappa}_{1}$ than to $-\bm{\kappa}_{2}$ is mapped to
   the spin-$\downarrow$ state with free-space momentum $\mathbf{q} =
   \mathbf{k}-\bm{\kappa}_{1}$; if $\mathbf{k}$ is closer to
   $-\bm{\kappa}_{2}$, the Bloch state is mapped to the spin-$\uparrow$ state
   with free-space momentum $\mathbf{q} = \mathbf{k}+\bm{\kappa}_{2}$.
   This construction is illustrated in Fig.~\ref{fig:bandmapping}.

For a completely filled lowest band the occupied states appear then as a hexagram (superimposed triangles for spins $\uparrow$ and $\downarrow$) after band mapping as shown in Fig.~\ref{fig:bandmapping}(b). A spin-resolved image of the cloud of atoms after band mapping, obtained either by appropriate detuning of imaging lasers or with a Stern-Gerlach filter, therefore allows for the reconstruction of the occupation numbers $n_{\mathbf{k}}$.
Finally, we note that signatures of the change in Fermi surface
  topology will also appear in the Hall response~\cite{spielmanhall},
  since the Berry curvature of the lowest band of the optical flux
  lattice is nonuniform~\cite{pricecooper}, so the Hall coefficient is
  sensitive to the distribution $n_{\mathbf{k}}$.

We thank Z. Hadzibabic, M. Fischer, G. Conduit and J. Levinsen for useful discussions. The work has been supported by EPSRC Grants  No. EP/I010580/1 and No. EP/F032773/1.
\maketitle


\begin{thebibliography}{10}

\bibitem{Duine:2005lh}
R.~A. Duine and A.~H. MacDonald, Phys. Rev. Lett. {\bf 95},  230403  (2005).

\bibitem{Conduit:2009qf}
G.~J. Conduit and B.~D. Simons, Phys. Rev. Lett. {\bf 103},  200403  (2009).

\bibitem{Jo:2009fu}
G.-B. Jo {\it et~al.}, Science {\bf 325},  1521  (2009).

\bibitem{Pilati:2010nx}
S. Pilati, G. Bertaina, S. Giorgini, and M. Troyer, Phys. Rev. Lett. {\bf 105},
   030405  (2010).

\bibitem{Chang:2010cr}
S.-Y. Chang, M. Randeria, and N. Trivedi, Proc. Natl. Acad. Sci. {\bf 108},  51
   (2011).

\bibitem{Pekker:2011oq}
D. Pekker {\it et~al.}, Phys. Rev. Lett. {\bf 106},  050402  (2011).

\bibitem{Sanner:2011kl}
C. Sanner {\it et~al.}, Phys. Rev. Lett. {\bf 108},  240404  (2012).

\bibitem{Fradkin:2010hc}
E. Fradkin {\it et~al.}, Annu. Rev. Condens. Matter {\bf 1},  153  (2010).

\bibitem{Lin:2009qf}
Y.~J. Lin {\it et~al.}, Nature {\bf 462},  628  (2009).

\bibitem{Wang:2012ve}
P. Wang {\it et~al.}, Phys. Rev. Lett. {\bf 109},  095301  (2012).

\bibitem{zwierlein}
L.~W. Cheuk {\it et~al.}, Phys. Rev. Lett. {\bf 109},  095302  (2012).

\bibitem{Williams:2012kx}
R.~A. Williams {\it et~al.}, Science {\bf 335},  314  (2012).

\bibitem{Cooper:2011kl}
N.~R. Cooper, Phys. Rev. Lett. {\bf 106},  175301  (2011).

\bibitem{Cooper:2011cr}
N.~R. Cooper and J. Dalibard, Europhys. Lett. {\bf 95},  66004  (2011).

\bibitem{Cui:2012tg}
X. Cui, Phys. Rev. A {\bf 85},  022705  (2012).

\bibitem{Tang:2011il}
E. Tang, J.-W. Mei, and X.-G. Wen, Phys. Rev. Lett. {\bf 106},  236802  (2011).

\bibitem{footnote}
We discretize momentum space with up to $54 \times 54$ grid points, which is sufficient to remove any finite size effects from the results presented.

\bibitem{footnote2}
When $|a_s|/l_z \gtrsim 1$ the Hamiltonian Eq. \eqref{eq:hlb} is still valid as an effective Hamiltonian with the bare interaction strength replaced with a renormalized coupling constant.

\bibitem{conduit}
G.~J. Conduit, Phys. Rev. A {\bf 82},  043604  (2010).

\bibitem{footnote3}
We have verified that the chemical potential lies below the minimum of the first excited band for the parameters considered here, so only the lowest Bloch band is populated.

\bibitem{petrov}
D.~S. Petrov, Phys. Rev. A {\bf 67},  010703  (2003).

\bibitem{ketterle}
C. Sanner {\it et~al.}, Phys. Rev. Lett. {\bf 108},  240404  (2012).

\bibitem{Pricoupenko:2007fk}
L. Pricoupenko and M. Olshanii, J. Phys. B {\bf 40},  2065  (2007).

\bibitem{Koehl2012}
M. Koschorreck {\it et~al.}, Nature {\bf 485},  619  (2012).
  
\bibitem{Ngampruetikorn:2011zr}
V. Ngampruetikorn, J. Levinsen, and M.~M. Parish, Europhys. Lett.
  {\bf 98},  30005  (2012).

\bibitem{Yamase:2000kx}
H. Yamase and H. Kohno, J. Phys. Soc. Jpn. {\bf 69}, 332 (2000); {\bf 69}, 2151 (2000).

\bibitem{Halboth:2000kx}
C.~J. Halboth and W. Metzner, Phys. Rev. Lett. {\bf 85},  5162  (2000).

\bibitem{Yamase:2005uq}
H. Yamase, V. Oganesyan, and W. Metzner, Phys. Rev. B {\bf 72},  035114
  (2005).

\bibitem{Jordens:2010vn}
R. J\"ordens {\it et~al.}, Phys. Rev. Lett. {\bf 104},  180401  (2010).

\bibitem{Zhai:2009}
H. Zhai, Phys. Rev. A {\bf 80}, 051605 (R) (2009).

\bibitem{Ma:2012}
     P.~N. Ma, S. Pilati, M. Troyer, and X. Dai, Nature Phys. {\bf 8}, 601 (2012).
       
\bibitem{dassarmapinczuk}
{\em Perspectives in Quantum Hall Effects: Novel Quantum Liquids in
  Low-Dimensional Semiconductor Structures}, edited by S. Das~Sarma and A.
  Pinczuk (Wiley, New York, 1997).

\bibitem{Saiga:2006fk}
Y. Saiga and M. Oshikawa, Phys. Rev. Lett. {\bf 96},  036406  (2006).

\bibitem{Katsura:2010}
H. Katsura {\it et~al.},  Europhys. Lett. {\bf 91} 57007 (2010).

\bibitem{Ciurylo:2005qa}
R. Ciury\l{}o, E. Tiesinga, and P.~S. Julienne, Phys. Rev. A {\bf 71},  030701
  (2005).

\bibitem{Reichenbach:2009mi}
I. Reichenbach, P.~S. Julienne, and I.~H. Deutsch, Phys. Rev. A {\bf 80},
   020701  (2009).

\bibitem{Hung:2011fu}
C.-L. Hung, X. Zhang, N. Gemelke, and C. Chin, Nature {\bf 470},  236  (2011).

\bibitem{Greiner:2001ye}
M. Greiner {\it et~al.}, Phys. Rev. Lett. {\bf 87},  160405  (2001).

\bibitem{Kohl:2005qo}
M. K\"ohl {\it et~al.}, Phys. Rev. Lett. {\bf 94},  080403  (2005).

\bibitem{spielmanhall}
L.~J. LeBlanc {\it et~al.}, Proc. Natl. Acad. Sci. {\bf 109},  10811  (2012).

\bibitem{pricecooper}
H.~M. Price and N.~R. Cooper, Phys. Rev. A {\bf 85},  033620  (2012).

\end{thebibliography}
\end{document}